\documentclass{appolb}
\usepackage{graphicx}

\begin{document}
\title{First estimation of the fission dynamics of the spectator created in
heavy-ion collisions
\thanks{}%
}
\author{
       K. Mazurek$^1$,  A. Szczurek$^1$,  P.N.Nadtochy$^2$ 
\address{$^1$Institute of Nuclear Physics PAN, ul. Radzikowskiego 152, Pl-31342
Krak\'ow, Poland\\
           $^2$Omsk State Technical University, Mira prospekt 11, Omsk, 644050, Russia\\
          }
           }
\maketitle
\begin{abstract}
In peripheral high-energy heavy-ion collisions only parts of colliding nuclei interact
leading to the production of a fireball. The remnants of such nuclei are
called spectators.
We estimated the excitation energy of nuclear remnants as a function of
impact parameter.  Their excitation energy is of the order of 100 MeV. 
The dynamical evolution of hot nuclei is described by solving a set of Langevin equations in four-dimensional collective
coordinate space. The range of nuclear masses and excitation energies
suits very well the ability of our model. Thus for the first time we investigate dynamically the fission and evaporation channels in de-excitation of
the spectators produced in heavy-ions collision. 
\end{abstract}
  
\section{Introduction}
The physics of high-energy heavy-ion collisions develops very effectively since many years. At CERN e.g. $^{208}$Pb+$^{208}$Pb collisions were considered at SPS and LHC. Two lead nuclei flying with velocities comparable with light velocity
produce the fireball in the place of collision, but depending on the centrality of the collision the Pb parts which are not 
participating in the collision (called ''spectators'') follow their initial path and deexcites later on. There are many experimental and 
theoretical \cite{teaney:2001,florkowski:2010} 
investigations of the processes connected to the fireball -- a piece of quark-gluon plasma. The density of the matter in the contact area is so high that the quark-gluon plasma is created 
and the high-energy particles such as pions, kaons, and others emerge in the hadronisation process. This part of the reaction requires dedicated models and methods that are beyond the scope of the present discussion.

The interesting issue is the behavior of the spectators produced as the remnants of the collision \cite{appelshauser:1998a}. There are not so many experimental 
observables published up to now, which can help with constraining the physics of spectator evolution. The fast collision
of two spherical ions provides exotic, for nuclear physics, shapes of the remnants which are strongly dependent on the impact parameter. The deformation energy of the produced system
turns into excitation energy. Thus we treat the spectator as an excited nucleus, which can deexcite by shape changes or
emission of light particles and/or $\gamma$-rays. This allows to apply the stochastic approach based on the solving transport equations 
in collective coordinate space \cite{kramers:1940,krappe:1995,abe:1996,frobrich:1998}.

\section{Theoretical method}

The stochastic approach assumes that the evolution of the nucleus in collective coordinate space is
similar to the Brownian motion of particles which interact stochastically 
involving a large number of internal degrees of freedom, which constitutes the surrounding "heat bath". 
The motion of the nucleus is slowed down by a friction, considered here as the hydrodynamical friction force.
The transport coefficients are derived from the random force
averaged over a time larger than the collision time scale between collective and internal
variables. The Gaussian white noise is taken for the random part, producing fluctuations of the physical observables
such as the mass/charge distribution of the fission fragments or evaporation residua.
The coupled Langevin equations describing the fission reaction have the form:
\begin{eqnarray}
\label{eq2}
\frac{d q_{i}}{d t}&=&\mu_{ij}p_{j},   \\
\frac{d p_{i}}{d t}&=&
- \frac{1}{2}p_{j}p_{k}\frac{\partial \mu_{jk}}{\partial q_{i}}
- \left( \frac{\partial F}{\partial q_i} \right)_T
- \gamma_{ij}\mu_{jk}p_{k}+\theta_{ij}\xi_{j}\left(t\right),
\nonumber
\end{eqnarray}
where ${\bf q}$ are the collective coordinates, ${\bf p}$
are the conjugate momenta, $F({\bf q},K)=V({\bf q},K) - a({\bf q}) T^2$ 
is the Helmholtz free energy, $V({\bf q})$ is the
potential energy, $m_{ij}({\bf q})$
($\|\mu_{ij}\|=\|m_{ij}\|^{-1}$) is the tensor of inertia, and
$\gamma_{ij}({\bf q})$ is the friction tensor. The $\xi_j\left(t\right)$ 
is a random variable satisfying the relations
\begin{eqnarray}
<\xi_{i}>&=&0, \nonumber \\
<\xi_{i}(t_{1})\xi_{j}(t_{2})>&=&2\delta_{ij}\delta(t_{1}-t_{2}).
\end{eqnarray}
Thus, a Markovian approximation is assumed to be valid.
The strength of the random force $\theta_{ij}$ is given by the Einstein 
relation $\sum \theta_{ik}\theta_{kj} = T\gamma_{ij}$. The temperature of
the "heat bath" $T$ is determined by the Fermi-gas model
formula $T=(E_{\rm{int}}/a)^{1/2}$, where $E_{\rm{int}}$ is 
the internal excitation energy of the nucleus and $a({\bf q})$ is 
the level-density parameter.

The energy conservation law is used
in the form 
\begin{eqnarray}
E^{*}=E_{\rm{int}}+E_{\rm{coll}}+V+ E_{\rm{evap}}(t)
\end{eqnarray}
at each time step during a random walk along the Langevin trajectory in the collective coordinate space. 
The total excitation energy of the nucleus is $E^{*}$ and the collective kinetic energy is taken as:
$E_{\rm{coll}}= 0.5 \sum \mu_{ij} p_i p_j$. The energy carried away by the evaporated
particles by the time $t$ is marked as $E_{\rm{evap}}(t)$. 

The collective coordinate space is constructed with three deformation parameters: elongation of the nucleus $(q_1)$, 
its neck $(q_2)$ and octupole shape $(q_3)$
based on the ''funny hills'' parametrization proposed initially by Brack~\cite{brack:1972} and the forth degree of freedom is the
projection of the angular momentum vector on the fission axis. This four collective coordinates allow to describe rich ensemble of shapes and aditinally it allows to orient the nucleus in the laboratory space, because $K$-is the projection of the spin on the fission axis.

The 4D Langevin method describes the behavior of the hot nucleus in the reference frame connected with the center-of-mass of each nucleus, 
thus in the case of the fast moving systems in the laboratory frame, additional transformations are necessary as the velocities of nuclei 
are in relativistic regime.

The initial condition to start evolution of the nucleus are usually the mass and charge of the beam and target nuclei, 
their excitation energy and maximal angular momentum, which can be produced in the reaction.

Our idea of investigating the spectator with 4D Langevin method is very recent. There are no direct measurements that could 
identify any fission/evaporation reactions occurring in the remnant of heavy-ions collisions, but new developments in the 
detection techniques \cite{staszewski:2016}
could make possible the investigations of new effects accompanying the main reaction.

The estimation of the shape of the spectator just after the ion collision is the gateway to introduce the fission dynamics.
We propose three possible scenarios of the non-central collisions. Let us assume that the beam and target nuclei have the spherical forms in their rest frame.
When they fly in opposite directions, collide and they graze each other. The central part produces the quark-gluon plasma and all kinds of high-energy particle phenomena.
Remnants could have extremely exotic forms and sometimes it is assumed that they are a collection of free nucleons. Here we discuss 
the following scenarios of initial shape production: ''sphere-plane'' means the sphere cut by the plane at various impact parameters; ''sphere-sphere'' 
gives the shape of the spectator which is sphere but with missing spherical cap; ''sphere-cylinder'' is the globe shot by the bullet.

These three scenarios gives shapes, very strange for nuclear physics. These exotic deformations are impossible to describe with any standard nuclear shape parameterizations. Thus we make an assumption that the time 
to change nuclear deformation from initial (complicated) to spherical, is small compared to the time-scale of the rest of the dynamics. 
This rough assumption will be investigated in the future in detail. The difference in the surface energy for spherical and 
deformed nucleus give the preliminary excitation energy of the spectator.

The potential energy of the nucleus is obtained with the Lublin-Strasbourg Drop formula \cite{pomorski:2003,dudek:2004}:
  \begin{eqnarray}
      E_{lsd} (Z,N;q)
      &=&E_{vol}+E_{surf}+E_{curv}+E_{_{Coul}}\\&=&
      b_{vol} (1-\kappa_{vol} I^2) \,A 
      +
      b_{surf} (1-\kappa_{surf} I^2)\, A^{2/3}B_{surf} (q)\nonumber\\ 
      &+&
      b_{curv} (1-\kappa_{curv} I^2)\, A^{1/3}B_{curv} (q)
      +
      \frac{3}{5}e^2 \frac{Z^2}{r^{ch}_{0}A^{1/3}} B_{Coul} (q),\quad
  \end{eqnarray}
where $ I =(N-Z)/A$. The geometric factors: $B_{surf}$, $B_{curv}$ and $B_{Coul}$ are defined as: 

\begin{eqnarray}
  B_{surf}=\frac{S(def)}{S(0)},\quad B_{curv}=\frac{C(def)}{C(0)}\quad \mathrm{ and }\quad B_{Coul}=\frac{E_{Coul}(def)}{E_{Coul}(0)}
\end{eqnarray}
that are the ratios between the surface($S$)/curvature($C$)/Coulomb energy ($E_{Coul}$) of the deformed and spherical nucleus. The details are presented e.g. in \cite{pomorski:2003}.
Thus for each of the scenarios, the surface and the volume
of the spectator were calculated as functions of the impact parameter. The volume of the spectator and
its mass were obtained using the geometrical properties. Assuming $N/Z_{Pb}=N/Z_{spectator}$ also the charge was assessed. 
The deformation together with the mass and charge gives the surface energy estimation.

\section{Results}
\begin{figure}%
\centering
\resizebox{0.8\textwidth}{!}{%
\includegraphics{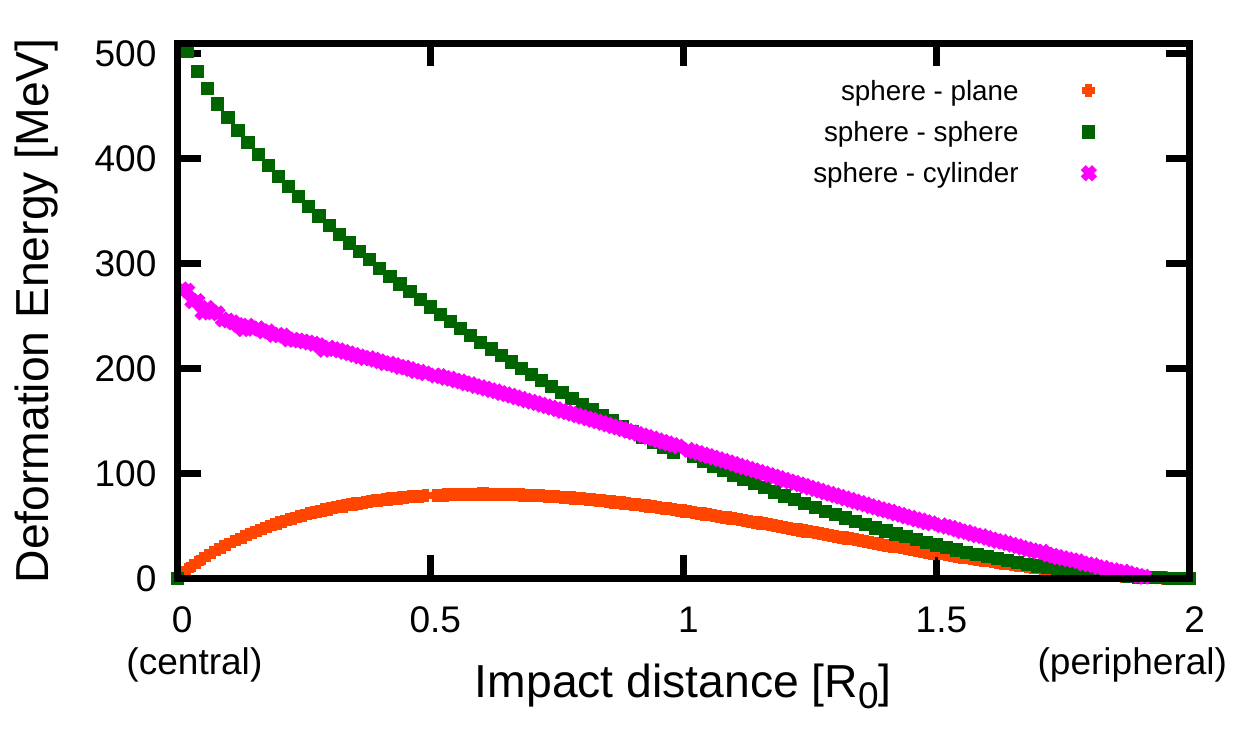}}
\caption{(Color on line) The deformation energy for three scenarios: ''sphere-sphere'', ''sphere-plane'' and ''sphere-cylinder''. The impact distance is normalized to the radius of the spherical lead nucleus.}
\label{fig_01}%
\end{figure}

\begin{figure}%
\centering
\resizebox{0.8\textwidth}{!}{%
\includegraphics{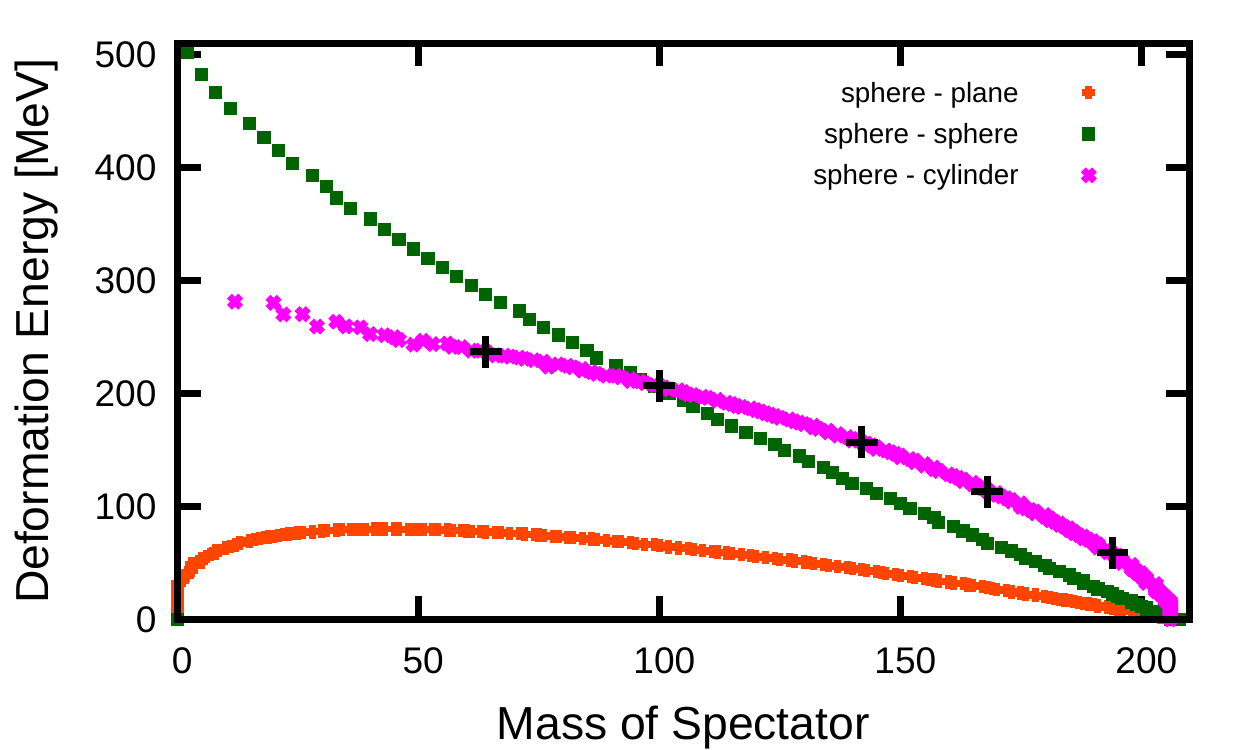}}
\caption{(Color on line) The deformation energy versus the spectator mass estimated from its volume. The crosses represent nuclei taken for dynamical evaluation.}
\label{fig_02}%
\end{figure}

The surface energy of the remnant just after the collision reduced by the energy of the spherical nucleus gives the deformation energy
which is presented in Fig.~\ref{fig_01}. The deformation energy depends on the initial condition. The highest deformation energies
(almost 500~MeV) are
predicted for the case of ''sphere-sphere'' grazing as the shapes are more curved than in other cases. The smallest energies are obtained for the
sphere cut by the plane as the forms are less curved and the surface of the deformed nucleus is closer to a spherical shape.
The impact parameter is good observable to discuss the central-peripheral collision, as it is presented in Fig.~\ref{fig_01}, but for fission dynamics calculation the mass and charge of the spectators are necessary.
Thus in Fig.~\ref{fig_02} the deformation energy is shown as a function of the mass of the remnant. The deformation energy has the main contribution to the excitation energy of the spectator. 
The black crosses mark the nuclei 
taken in further calculations.
The 4D Langevin method is dedicated to describe the fission dynamics of the hot nuclei at excitation energies $E^{\star}$=50--250~MeV and the driving forces are derived macroscopically.
The excitation energy was taken as the deformation energy from the scenario: ''sphere-cylinder'' as the energy range fits exactly limitation of our model.
From phenomenological point of view this model is also the most realistic.
\begin{figure}%
\centering
\resizebox{1.1\textwidth}{!}{%
\includegraphics{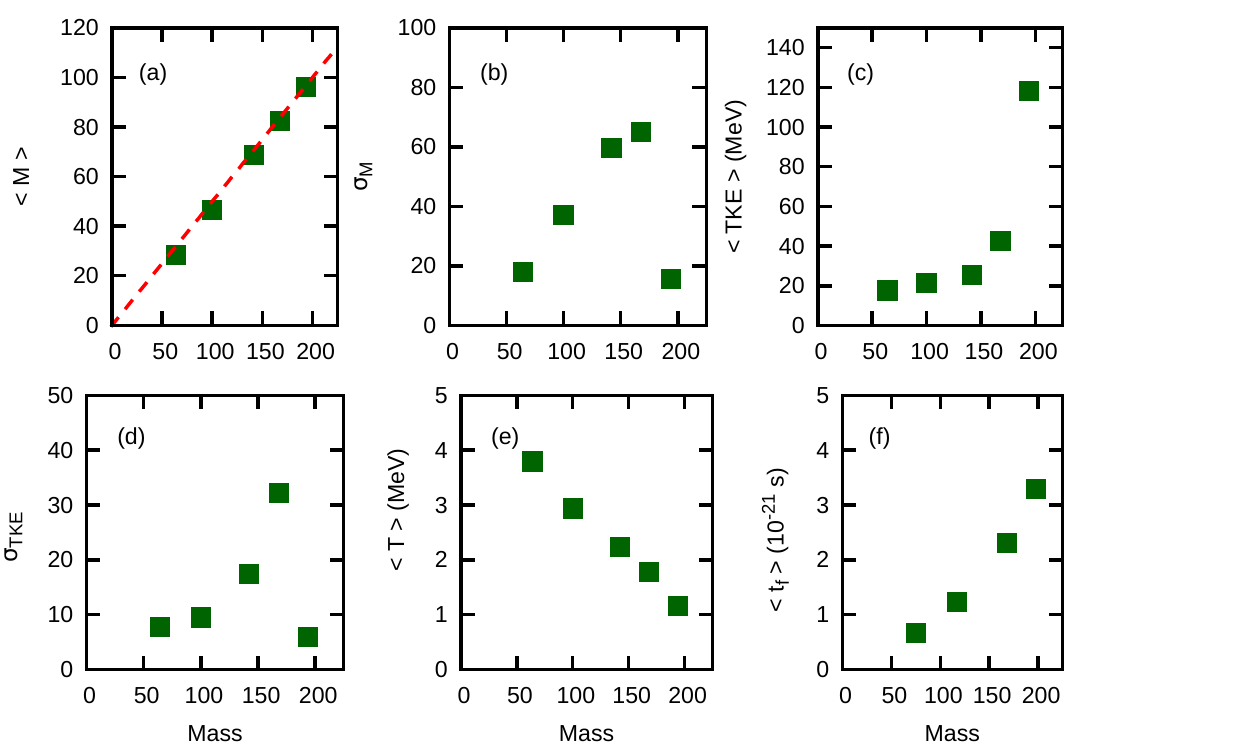}}
\caption{(Color on line) The average values of the mass (a), total kinetic energy (c), temperature (e) and fission time (f) resulting from initial systems with 
M=64, 100, 142, 168 and 194. The widths of the mass (b) and TKE distribution (d) are also displayed. The dashed line in panel (a) shows the mean mass of the fission fragments assuming symmetric scission of the nucleus. }
\label{fig_03}%
\end{figure}
Five nuclei are chosen arbitrarily for the fission dynamics studies: $^{64}$Mn, $^{100}$Y, $^{142}$Ba, $^{168}$Dy and $^{194}$Os.
The maximal angular momentum was taken as 2~$\hbar$ just to avoid possible technical problems. For each nucleus around 300 000 trajectories were 
calculated and the results are presented in Fig.~\ref{fig_03}. The panel (a) compares the mean mass coming from an assumption 
of the symmetric division of the fissioning nucleus (red dashed line) and centroid of the mass distribution whose width is shown in panel (b). 
The medium mass system demonstrates a slight deviation from the static assumption as during the path to fission various particles (protons. neutrons, deuteron, $\alpha$-particles) 
are emitted. Looking at panel (c) for the Total Kinetic Energy (TKE) of the fission fragments, it is visible that almost linear decrease of 
the excitation energy (Fig.~\ref{fig_02}) provides the exponential increase of the average TKE as the system is more and more heavy.
The effects of inertia and/or the excitation energy is also visible in panel (f) for the increased fission time.

\section{Summary}

The preliminary results discussed here show a possibility of investigating the fission of spectators produced in ultrarelativistic heavy-ions collisions.
The three scenarios of initial shape estimation for the remnants of the Pb-Pb reaction have been considered to estimate the excitation energies available for various impact parameters. Several observables have been extracted from the state-of-art Langevin calculations for fission dynamics. The average values and widths of mass, charge and total kinetic energy distribution of the fission fragments as well as mean values of the temperature and fission times have been shown for the first time.

\vskip 1cm
{\bf Acknowledgements}
\\
The work was partially supported by the Polish
National Science Centre under Contract No. 2013/08/M/ST2/00257 (LEA COPIGAL) (Project No.~18) and IN2P3-COPIN (Project No.~12-145, 09-146), and by the Russian Foundation for Basic Research (Project No.~13-02-00168). 


\end{document}